\documentclass[prd,aps,twocolumn,nofootinbib]{revtex4}

\newcommand{\bear}{\begin{array}}  \newcommand{\eear}{\end{array}}
\newcommand{\bea}{\begin{eqnarray}}  \newcommand{\eea}{\end{eqnarray}}
\newcommand{\beq}{\begin{equation}}  \newcommand{\eeq}{\end{equation}}
\newcommand{\bef}{\begin{figure}}  \newcommand{\eef}{\end{figure}}
\newcommand{\bec}{\begin{center}}  \newcommand{\eec}{\end{center}}

\newcommand{\la}{\left\langle} \newcommand{\ra}{\right\rangle}

\newcommand{\ds}{\displaystyle}
\newcommand{\be}{\begin{eqnarray}}
\newcommand{\ee}{\end{eqnarray}}

\def\EQ#1{Eq.~(\ref{#1})}

\begin{document}

\title{511 keV line from Q balls in the Galactic Center}

\author{Shinta Kasuya$^a$ and Fuminobu Takahashi$^{b,c}$}

\affiliation{
$^a$ Department of Information Science,
     Kanagawa University, Kanagawa 259-1293, Japan\\
$^b$ Institute for Cosmic Ray Research,
     University of Tokyo, Chiba 277-8582, Japan\\
$^c$ Deutsches Elektronen-Synchrotron DESY, 22603 Hamburg, Germany}

\date{August, 2005}

\begin{abstract}
The 511 keV photons from the galactic center can be explained by
positrons produced through Q-ball decay.
In the scheme of gauge-mediated supersymmetry breaking, large Q balls
with lepton charge are necessarily long-lived. In particular, the
lifetime can be as long as (or even longer than) the age of the universe.  
If kinematically allowed, such large Q balls decay into
positrons, which eventually annihilate with electrons into 511 keV
photons.  Our scenario is realized within the minimal supersymmetric
standard model in the inflationary universe, 
which is very plausible.
\end{abstract}


\maketitle

\setcounter{footnote}{1}
\renewcommand{\thefootnote}{\fnsymbol{footnote}}

\section{Introduction}
The 511 keV gamma-ray line emission from the galactic bulge was
observed by the spectrometer SPI on the International Gamma-Ray
Astrophysics Laboratory (INTEGRAL)
\cite{SPI}. With the unprecedented resolution, it revealed that
the emission comes from spread (or multiple point) sources, which
makes it very difficult to explain with traditional astrophysical
objects, such as neutron stars, black holes, radioactive nuclei from
supernovae, novae, red giants, and Wolf-Rayer stars, cosmic ray
interactions with the interstellar medium, pulsars or stellar flares,
type Ia supernovae, hypernovae, etc.

One of the alternative candidates for the 511 keV line was provided by
annihilating or decaying dark matter whose decay products contain
positrons \cite{DDM1,DDM2}.  In order to have the observed flux of the
511 keV line emission, decaying dark matter (DDM) scenario may be
characterized into the two extreme cases. One is that the DDM accounts
for the present matter density of the universe ($\rho_{ddm} \sim
\rho_m$), with extremely long lifetime ($\tau_{ddm} \gg t_0$). The other has 
lifetime as long as the age of the universe ($\tau_{ddm} \sim t_0$), while 
its abundance is very small ($\Omega_{ddm} \ll 1$). The observed flux of 
the 511 keV line emission is in general related to the abundance and the 
lifetime of the decaying particle as \cite{DDM2} 
\begin{equation}
\frac{\Phi_{511}}{10^{-3} /{\rm cm}^2/{\rm sec}} \simeq \Omega_{ddm} \left(\frac{\tau_{ddm}}{10^{27}~{\rm sec}}\right)^{-1} 
\left(\frac{m_{ddm}}{\rm MeV}\right)^{-1},
\label{eq:rel}
\end{equation}
where we assume spherically symmetric profile for dark matter density with  
$\rho \propto r^{-1.2}$.
In order to achieve this condition, the scenarios proposed thus far
need somewhat artificial and {\it ad hoc} assumptions about couplings,
and sometimes huge dilution by late entropy production.

In this article, we show instead more natural candidate for the source
of positrons in the galactic bulge: Q balls~\cite{Coleman} in the
minimal supersymmetric standard model (MSSM).
 Throughout this paper we assume the gauge-mediated supersymmetry
 (SUSY) breaking.  Our promising candidate is a large Q ball with the
 charge being the lepton number.
The large charge enables Q balls not only to be long-lived but also to
have small mass per unit charge.
Due to these features, we can successfully explain the origin of the
galactic positrons with relatively small energy ($ \lesssim O(10)$
MeV)~\cite{Beacom:2004pe}.

The organization of the article is as follows. In the next section, we
review several properties of two different types of Q balls in the
MSSM with gauge-mediated SUSY breaking.  In Sec.~\ref{sec:sec3}, we
will estimate how large the charge should be for the Q balls to be
enough long-lived.  The 511 keV gamma ray flux from the Q-ball decay
will be calculated in Sec.~\ref{sec:sec4}. Sec.~\ref{sec:sec5} is
devoted for the Q-ball formation processes and a possible way to
give the small enough abundance of the Q balls. Finally, we give 
conclusions in Sec.~\ref{sec:sec6}.

\section{Q balls in the gauge-mediated SUSY breaking}
\label{sec:sec2}
A Q ball is a nontopological soliton composed of a complex scalar
field $\Phi$, and it minimizes the energy of the scalar field with a
fixed $U(1)$ charge $Q$ \cite{Coleman}.  In MSSM, this charge is
actually some combination of baryon and lepton numbers. Since our aim
here is to explain the positron flux, we identify the $U(1)$ charge
with the lepton number.  Such Q balls are often called `L balls'.

MSSM contains many flat directions along which the scalar potential
vanishes at the level of renormalizable operators in the global SUSY
limit.  The flat directions generally consist of squarks, sleptons and
higgs \cite{FD}, and are fully classified in
Ref.~\cite{Gherghetta:1995dv}.  Here we consider the leptonic
directions such as $eLL$.  The scalar potential is lifted by the SUSY
breaking effects and presumably by nonrenormalizable operators in the
superpotential. Since the existence of the nonrenormalizable operators is irrelevant
to the following discussion, we will drop them for simplicity. In fact
they can be forbidden in the presence of the discrete gauge
symmetry. Also, for the moment, let us concentrate on the $U(1)$
conserving part of the scalar potential, since $U(1)$ violating
interactions are not important for determining properties of the Q
balls (however, see Ref.~\cite{Kawasaki:2005xc}).

In the gauge-mediated SUSY breaking, the scalar potential of the flat
direction is parabolic below the messenger scale $M_S$, while, above
that scale, the potential grows
logarithmically~\cite{deGouvea:1997tn}:
\begin{equation}
\label{eq:vgauge}
    V_{gauge}(\Phi) \sim \left\{ 
      \begin{array}{ll}
          m_{\phi}^2|\Phi|^2 & \quad (|\Phi| \ll M_S) \\
          \ds{M_F^4 \left(\log \frac{|\Phi|^2}{M_S^2} \right)^2}
          & \quad (|\Phi| \gg M_S) \\
      \end{array} \right.,
\end{equation}
where $m_{\phi}$ is a soft breaking mass $\sim O$(TeV), and $M_F$ and
$M_S$ are related to the $F$- and $A$- components of a gauge-singlet
chiral multiplet $S$ in the messenger sector as
\beq
M_F^4 \equiv \frac{g^2}{(4 \pi)^4} \la F_S \ra^2,
\,\,\,\,\, M_S \equiv \la S \ra.
\eeq
Here $g$ generically stands for the standard model gauge coupling.
The masses of the sparticles in the visible sector is given by $m \sim
g^2 \Lambda_{mess}/(4\pi)^2$, where $\Lambda_{mess}= \la F_S \ra / \la
S \ra \sim 10^5$ GeV.
Noting that $\la F_S \ra^{1/2} \sim \la S \ra$ should not be necessarily
satisfied, the allowed range for $M_F$ is
\begin{equation}
\label{MF}
10^3 {\rm GeV} \lesssim M_F \lesssim \frac{g^{1/2}}{4\pi} \sqrt{m_{3/2} M_P},
\end{equation}
where $m_{3/2}$ is the gravitino mass and $M_P=2.4 \times 10^{18}$ GeV
is the reduced Planck mass.

In addition to the gauge-mediated effects, there is always the
gravity, which also mediates the SUSY breaking, leading to
\begin{equation}
V_{grav}(\Phi) = m_{3/2}^2 \left[ 1+ K \log\left( \frac{|\Phi|^2}{M_P^2} \right) \right] |\Phi|^2.
\end{equation}
Here we include the one-loop corrections to the mass term. For most
flat directions, the numerical coefficient $K$ is negative and varies
between $-0.1$ and $-0.01$.  The scalar potential is given by the sum
of $V_{gauge}$ and $V_{grav}$.  Since the gravitino mass is much
smaller than the weak scale, $V_{grav}$ dominates only at large field
amplitudes.

The Q-ball solution exists if and only if $V(\phi)/\phi^2$ has a
minimum at $\phi \ne 0$, where $\phi \equiv \sqrt{2} |\Phi|$ is the
radial part of $\Phi$.  In other words, the effective potential must
be shallower than $\phi^2$.  The gauge-mediation potential
$V_{gauge}$, as well as the gravity-mediation potential $V_{grav}$
with negative $K$, satisfies this criterion.  Therefore, there are two
types of the Q ball in the gauge-mediated SUSY breaking models,
depending on the value of $\phi$.  If the field amplitude is smaller
than $\phi_{eq} \sim M_F^2/m_{3/2}$, the gauge-mediation potential
dominates. We call this the gauge-mediation (GM) type Q ball.  Its
typical charge is related to the field amplitude at which the flat
direction $\Phi$ starts to oscillate~\footnote{
We describe the dynamics of the flat direction $\Phi$ and Q-ball formation in
Sec.~\ref{sec:sec5}.}
by \cite{KasuyaKawasaki}
\beq
Q \simeq \beta_{gauge} \left(\frac{\phi_{osc}}{M_F}\right)^4,
\eeq
where $\beta_{gauge} \approx 6 \times 10^{-4}$.  The mass and the size
of the GM-type Q ball are, respectively,
\begin{equation}
M_Q \sim M_F Q^{3/4}, \quad R_Q \sim M_F^{-1} Q^{1/4}.
\end{equation}
The mass per unit charge, $m_Q = M_Q /Q$, can be very small for large
enough $Q$. If $m_Q$ exceeds the mass of some particles carrying
lepton number, the L ball can decay into those particles. Thus, in
contrast to the Q balls having the baryon number, L balls can always
decay into neutrinos, unless $Q$ is so large to render $m_Q$ smaller
than the neutrino mass.

On the other hand, if the field amplitude is larger than $\phi_{eq}$,
the gravity-mediation potential dominates, and `new-type' Q balls are
generated \cite{Newtype}.  The typical charge is
\cite{KasuyaKawasaki,Newtype}
\beq
\label{Qgrav}
Q \simeq \beta_{grav} \left(\frac{\phi_{osc}}{m_{3/2}}\right)^2,
\eeq
where $\beta_{grav} \approx 6 \times 10^{-3}$. The mass and the size are
\begin{equation}
M_Q \simeq m_{3/2} Q, \quad R_Q \simeq  m_{3/2}^{-1} |K|^{-1/2},
\end{equation}
respectively. The mass per unit charge is now independent of $Q$, and
given by the gravitino mass: $m_Q \sim m_{3/2}$. The decay particle
species now depend on the gravitino mass. To produce positrons in the 
galactic center, $m_{3/2}$ should be larger than the positron mass, 
$511$~keV.  Since there is a variety of the gauge-mediated SUSY breaking
scenarios with the gravitino mass varying from $O$(GeV) well down to
$O$(keV), such a requirement can be easily satisfied.

\section{Q-ball decay and positron production}
\label{sec:sec3}
As mentioned in the previous section, we are interested in the
positron production from L balls.  The decay products generally
contain neutrinos, charged leptons, and their antiparticles. \footnote{
Branching ratio to photons can be extremely suppressed due to the very 
small left-right slepton mixings.}
In our case, the mass per unit lepton number, $m_Q$, should be larger than
$511$ keV, in order to produce positrons. In this section let us
derive a condition on each type of Q-balls to have lifetimes longer
than the present age of the universe, with $m_Q \sim $MeV fixed.

The Q-ball decay takes place only through its surface because of the
Pauli blocking and the large effective masses of the coupled particles
well inside the Q ball.  In the case of L balls, the decay proceeds
via gaugino exchange into a pair of leptons \cite{Kawasaki:2002hq}.
The charge decreasing rate is constrained from above
\cite{Cohen:1986ct},
\begin{equation}
-\frac{dQ}{dt} \leq \frac{m_Q^3}{192\pi^2} {\cal A},
\end{equation}
where ${\cal A}=4\pi R_Q^2$ is the surface area of the Q ball. The
actual decay rate of the L balls should be close to this upper bound.
We consider the two types of Q balls one by one.

\subsection{GM-type Q balls}
The decay rate of the GM-type Q ball reads
\begin{equation}
\Gamma_Q = -\frac{1}{Q} \frac{dQ}{dt}  \sim \frac{M_F}{48\pi Q^{5/4}}.
\end{equation}
As mentioned in Introduction, the life time of the L ball must be as
long as, or longer than, the present age of the universe, $t_0 \sim$
13 Gyr. To this end, a huge charge is necessary:
\begin{equation}
\label{Q1}
Q \gtrsim 10^{36} \left(\frac{M_F}{10^6~{\rm GeV}}\right)^{4/5}.
\end{equation}
In addition, the mass per unit charge must be around $O$(MeV), in
order for the L balls to emit positrons:
\begin{equation}
Q  \sim  \left(\frac{M_F}{\rm MeV}\right)^4
\sim 10^{36} \left(\frac{M_F}{10^6~{\rm GeV}}\right)^4.
\end{equation}
These two conditions meet when $Q \sim 10^{36}$ and $M_F \sim 10^6$
GeV. Notice that the value coincides with the upper limit of $M_F$ in
Eq.(\ref{MF}).

On the other hand, $\phi_{osc}$ should be smaller than $\phi_{eq}$ to
produce GM-type Q-balls:
\begin{equation}
\phi_{osc} \leq \phi_{eq} \sim \frac{M_F^2}{m_{3/2}} \lesssim 10^{-3} M_P,
\end{equation}
where we used Eq.(\ref{MF}) in the last inequality. Thus, the charge
of the Q ball is bounded from above~\footnote{%
 If $K$ is positive,
another type of Q balls dubbed ``delayed-type'' Q-balls are
formed~\cite{KKT}.  The properties of the delayed-type Q-balls are
same as the GM-type Q-balls, except for $\beta_{gauge} \sim 1$.  But
even in this case, the maximum charge is marginally too small.}
\begin{equation}
Q \lesssim \beta_{gauge} \left(\frac{\phi_{eq}}{M_F}\right)^4 \lesssim 6\times 10^{32},
\end{equation}
where we have substituted $M_F \sim 10^6$ GeV.  It is therefore
impossible to obtain large enough charge of $10^{36}$, forcing us to
conclude that the GM-type of the Q ball cannot account for the
observed 511 keV gamma ray, because of its too short lifetime.

\subsection{New-type Q balls}
For the new-type Q ball, the decay rate is given by
\begin{equation}
\Gamma_Q \sim \frac{m_{3/2}}{48 \pi |K| Q}.
\end{equation}
In order for the new-type Q balls to live longer than the age of the
universe, its charge should satisfy
\begin{equation}
Q \gtrsim 4 \times 10^{37} \left(\frac{0.1}{|K|}\right) \left(\frac{m_{3/2}}{\rm MeV}\right).
\end{equation}
Equivalently, from Eq.(\ref{Qgrav}), the field amplitude at which the
flat direction starts to oscillate is bounded below,
\beq
\phi_{osc} \gtrsim  10^{17} {\rm GeV} \left(\frac{0.1}{|K|}\right)^{\frac{1}{2}}
 \left(\frac{m_{3/2}}{\rm MeV}\right)^{\frac{3}{2}}.
\eeq
Such a large field value can be naturally attained if
nonrenormalizable terms in the superpotential are forbidden by some
symmetry.  In the following argument, let us concentrate on
the new-type Q-ball as a possible source for the positrons in the
galactic bulge.

\section{Positrons flux from Galactic center}
\label{sec:sec4}
The relation between the 511 keV gamma-ray flux and the properties
(i.e., lifetime, mass, and abundance) of the decaying dark matter is
given by \EQ{eq:rel}.  If the charge $Q$ is as large as $10^{47}$, the
new-type Q ball has lifetime as long as $10^{27}$ sec and accounts for
most of the dark matter. However, the maximal possible charge of the
new-type Q ball is attained when $\phi_{osc} \sim M_P$:
\begin{equation}
Q_{max} = \beta_{grav} \left(\frac{M_P}{m_{3/2}}\right)^2 
\sim 4 \times 10^{40} \left(\frac{m_{3/2}}{\rm MeV}\right)^{-2},
\end{equation}
which is much smaller than $10^{47}$.
In other words, the lifetime is too short for the Q balls to be the dominant
component of the dark matter. Using the relation (\ref{eq:rel}), the range of the charge
and the abundance of the Q ball are
\bea
4 \times 10^{37} \lesssim &Q&  \lesssim 4 \times 10^{40}, \\
4 \times 10^{-10} \lesssim &\Omega_Q& \lesssim 4 \times 10^{-7},
\label{eq:qball_ab}
\eea
respectively.
This corresponds to the lifetime: $ 4 \times 10^{17} {\rm sec} \lesssim \tau_Q \lesssim 
4 \times 10^{20} {\rm sec}$.

While the charge in the above range can be easily realized, the
abundance of such large Q balls tends to be too large, unless we
introduce some mechanism to suppress it.  One of the simplest
solutions is to dilute them by generating large entropy at later
epoch; it can be realized by e.g., thermal inflation~\cite{thermal} or
unstable domain-wall network~\cite{Kawasaki:2004rx}. However, since
the large entropy production makes it difficult to generate enough
baryon asymmetry, it is desirable if there is another way to reduce
the Q-ball density. In fact, as discussed in the next section,
nonperturbative decay of the flat direction could serve as another
solution, which can be naturally implemented in our scheme.

\section{Dynamics of the flat direction}
\label{sec:sec5}
Here we first explain the dynamics of flat directions and subsequent 
Q-ball formation. Second, we briefly discuss a possible way to reduce the
Q-ball density via nonperturbative decay of the flat
direction.

\subsection{Q-ball formation}

During inflation the flat direction is assumed to develop a large
vacuum expectation value $\phi_0$. We further assume that $\phi_0$ is
larger than $\phi_{eq}$ so that  new-type Q balls are formed.  When the Hubble
parameter becomes comparable to the gravitino mass, the flat direction
starts to oscillate. At the same time, it is kicked into the phase
direction due to $U(1)$-violating interactions dubbed A-term. Without
nonrenormalizable interactions in the superpotential, it comes from
higher order lepton number violating terms in the K\"ahler potential:
\begin{equation}
V_A(\Phi) \sim 
\frac{m_{3/2}^2 \Phi^n}{M_*^{n-2}} 
+ {\rm h.c.},
\label{eq:kahlerA-term}
\end{equation}
where $M_*$ is a cut-off scale of this interaction.

Once the rotation starts, the flat direction feels spatial instabilities due to 
negative pressure; the potential of the flat direction is proportional 
to $|\Phi|^{2+2K}$ with $K<0$, and the pressure of the oscillating (rotating) scalar field in this
potential is $p=-|K|\rho/2 < 0$. These instabilities continue to grow large until the 
field deforms into Q balls \cite{Qball,KasuyaKawasaki} through I-ball formation \cite{Kasuya:2002zs}.

The detailed numerical simulations  \cite{KasuyaKawasaki} showed
 that almost all the charge
is absorbed into the Q balls. The typical charge of the Q balls 
depends on the initial amplitude of the field, $\phi_{0}$, 
which is roughly same as $\phi_{osc}$. See \EQ{Qgrav}.

Since the energy density of the Q balls at their production is 
$\rho_Q \sim m_{3/2}^2 \phi_{osc}^2$, only slightly smaller than the total energy
density of the universe, it will dominate the universe soon after 
the reheating.  Even if the reheating temperature is as low as MeV \cite{Kawasaki:1999na}, 
the Q balls will overclose the universe in the end. This can be seen by calculating 
the Q-ball-to-entropy ratio:
\beq
\frac{\rho_Q}{s} 
 \sim  10 ~{\rm MeV}  \left(\frac{T_{RH}}{10 {\rm MeV}}\right) \frac{\phi_{osc}^2}{M_P^2}
\eeq
To satisfy (\ref{eq:qball_ab}), the Q-ball energy density must be off at least by $10^{14}$.
Clearly some remedy is necessary.

\subsection{Rapid decay of the flat direction}

In the previous subsection, we did not include the coupling of the flat 
direction to  other fields (say $\chi$). In fact, the flat direction decays into $\chi$ particles 
very efficiently if the 
rotational orbit is eccentric enough. $\chi$ particles are created nonperturbatively
while the flat direction passes near the origin. In general, $\chi$ has further 
couplings to other particles, symbolically denoted by $X$. Through the interactions,
the created $\chi$ particles may decay into $X$ immediately after their production. 
Such mechanism is known as instant preheating \cite{Preheating}. 

In order to produce $\chi$ particles, the effective mass of $\chi$ must 
change nonadiabatically.  The particle production occurs each time $\Phi$ 
passes near the origin. Assuming the following
interaction\footnote{
Such interaction comes from the $D$-term potential.
Although lepton asymmetry cannot be transmitted via this interaction,
it is transmitted through, e.g., Yukawa-type interactions with higgsino or
gauginos. In the latter case, the efficiency of the $\chi$-production is
more or less the same as that considered in the text.},
\begin{equation}
\label{eq:int}
{\cal L}_{int} = g^2|\Phi|^2 |\chi|^2,
\end{equation}
the adiabiticity is violated if
\begin{equation}
\dot{\omega}_k \gtrsim \omega_k^2,
\end{equation}
where $\omega_k^2=\sqrt{k^2+g^2\phi^2(t)}$. This inequality is satisfied only in the vicinity 
of the origin; $\phi \lesssim \phi_*$ with  $\phi_* \sim (m_{3/2} \phi_{osc}/g)^{1/2}$.
For  $m_{3/2} \sim$ MeV, $\phi_{osc} \sim 10^{17}$ GeV and $g \sim 0.3$, we have 
$\phi_* \sim 10^7$ GeV.
The nonperturbative particle production via the interaction (\ref{eq:int})
thus occurs if the ratio of major and minor axes of the elliptical orbit, $\varepsilon$,
is smaller than $\varepsilon_c \equiv \phi_*/\phi_{osc} \sim 10^{-10}$.  

Let us assume that the flat direction acquires quantum fluctuations during inflation.
If the averaged value of $\varepsilon$ is slightly smaller than $\varepsilon_c$,  the flat 
direction rapidly decays in most parts of the entire universe. In the meantime, the Q-balls
with very large $Q$ are formed in those patches where $\varepsilon$ is larger than $\varepsilon_c$ due to fluctuations and the instant preheating does not occur~\footnote{
Notice that
the Q-ball formation will proceed if the coherence is extended over 
$|K|^{-1/2} m_{3/2}^{-1}$, which is $\sim 10$ times larger than the horizon at the beginning of 
the oscillations. }.
If this is the case, we can suppress the Q-ball density while keeping the charge of the Q balls
very large.

For the A-term given by  \EQ{eq:kahlerA-term}, $\varepsilon$ is 
\beq
\varepsilon \sim \theta \left(\frac{\phi_{osc}}{M_*} \right)^{n-2},
\eeq
where $\theta$ is the CP phase. Such small $\varepsilon \lesssim 10^{-10}$ 
can be realized, for example, when $n=6$, $M_* \sim M_P$, and $\theta \sim 10^{-5}$.
The quantum fluctuations must be so large that $\delta \varepsilon/\varepsilon\sim H_{I}/(\phi_{osc} \theta)$ is of order unity~\cite{KT}, where $H_I$ is the Hubble parameter 
during inflation. To this end, $H_I$ should be around $10^{12}$GeV for the exemplified 
values of $\phi_{osc}$ and $\theta$.

Lastly let us comment on the asymmetry of the flat direction. It is easy to see that 
$\delta \varepsilon/\varepsilon$ gives a rough estimate of the isocurvature fluctuations
in the lepton number of the flat direction concerned. Therefore it is difficult to associate 
the lepton asymmetry of the flat direction responsible for $511$keV line with the primordial 
baryon asymmetry. However, since our model does not require late-time entropy production, 
any successful baryogenesis, including the Affleck-Dine mechanism~\cite{FD} using
a flat direction compatible with the leptonic direction we considered, should work.

\section{Conclusions}
\label{sec:sec6}
We have shown that the annihilation of positrons from the Q-ball decay can explain the
observed 511 keV line emision from the galactic center observed by SPI/INTEGRAL. 
The setup is natural and minimal in the sense that we have used the MSSM Q balls in 
the gauge-mediated SUSY breaking. One of the advantages using Q balls is that they 
naturally have both the small mass per unit charge of $O({\rm MeV})$ and the lifetime 
longer than the present age of the universe. This distinctive feature makes our scenario 
appealing. In addition, the desired Q-ball abundance can be realized without resort to large 
entropy production; the nonperturbative decay processes drastically suppress the abundance.

Finally we comment on the dark matter.  
Since the present energy density of the L ball cannot account for the dark matter, 
$\Omega_m \sim 0.3$, we need other candidates.  Among all, here we propose Q ball 
dark matter, which is made from the different flat direction.  The best candidate is the 
GM-type B ball with $n=6$ $udd$ direction, which may co-exist. Since the tiny fraction
of the large L balls does not change the thermal history of the universe so much, the 
B-ball dark matter (and possibly related baryogenesis) is as the same in 
Ref.\cite{KasuyaKawasaki}.

\section*{Acknowledgments}
The authors are grateful to M. Kawasaki for useful discussion. 
The work of S.K. is supported by the Grant-in-Aid for Scientific Research from the
Ministry of Education, Science, Sports, and Culture of Japan, No.~17740156.
F.T.  would like to thank the Japan Society for Promotion of Science for financial support. 



\end{document}